\documentclass[twocolumn]{aastex63}

\usepackage{amssymb, amsfonts, amsmath,framed}

\usepackage{bm}
\expandafter\ifx\csname package@font\endcsname\relax\else
 \expandafter\expandafter
 \expandafter\usepackage
 \expandafter\expandafter
 \expandafter{\csname package@font\endcsname}
\fi
\expandafter\ifx\csname package@font\endcsname\relax\else
 \expandafter\expandafter
 \expandafter\usepackage
 \expandafter\expandafter
 \expandafter{\csname package@font\endcsname}
\fi
\def\bi{\begin{itemize}}
\def\ei{\end{itemize}}
\def\bq{\begin{equation}}
\def\eq{\end{equation}}
\def\bqy{\begin{eqnarray}}
\def\eqy{\end{eqnarray}}

\begin{document}
\title{Hard Synchrotron Spectra from Magnetically Dominated Plasma Turbulence}

\correspondingauthor{}
\email{luca.comisso@columbia.edu \\ es3808@columbia.edu \\ lsironi@astro.columbia.edu}

\author{Luca Comisso}
\affiliation{Department of Astronomy and Columbia Astrophysics Laboratory, Columbia University, New York, NY 10027, USA}

\author{Emanuele Sobacchi}
\affiliation{Department of Astronomy and Columbia Astrophysics Laboratory, Columbia University, New York, NY 10027, USA}

\author{Lorenzo Sironi}
\affiliation{Department of Astronomy and Columbia Astrophysics Laboratory, Columbia University, New York, NY 10027, USA}

\begin{abstract}
Synchrotron emission from astrophysical nonthermal sources usually assumes that the emitting particles are isotropic.
By means of large-scale  two- and three-dimensional particle-in-cell simulations, we demonstrate that the dissipation of magnetically dominated ($\sigma_0\gg1$) turbulence in pair plasmas leads to strongly anisotropic particle distributions. At Lorentz factors $\sim \sigma_0 \gamma_{th0}$ (here, $\gamma_{th0}$ is the initial Lorentz factor), the particle velocity is preferentially aligned with the local magnetic field; instead, the highest energy particles are preferentially oriented in the plane perpendicular to the field. This energy-dependent anisotropy leads to a synchrotron spectral flux $\nu F_\nu\propto \nu^s$ that is much harder than for  isotropic particles. Remarkably, for $\sigma_0\gg1$ we find that the angle-integrated spectral slope in the slow cooling regime is $s\sim 0.5-0.7$ for a wide range of turbulence fluctuations, $0.25\lesssim \delta B_{\rm rms0}^2/B_0^2\lesssim 4$, despite significant variations in the power-law energy spectrum of nonthermal particles. This is because weaker turbulence levels  imprint a stronger degree of anisotropy, thereby counteracting the effect of the steeper particle spectrum. The synchrotron spectral slope may be even harder, $s\gtrsim 0.7$, if the observer is in the plane perpendicular to the mean magnetic field. Our results are independent of domain size and dimensionality. Our findings may help explain the origin of hard synchrotron spectra of astrophysical nonthermal sources, most notably the radio spectrum of pulsar wind nebulae.
\end{abstract}


\section{Introduction}

Synchrotron emission from a nonthermal population of energetic particles is invoked to explain the radiative signature of a variety of high-energy astrophysical sources. The emitting particles are often assumed to be distributed according to a power law with a slope $p$ in energy, $dN/d \gamma\propto \gamma^{-p}$, and their velocity to be isotropically oriented with respect to the local magnetic field. Under these assumptions, the synchrotron energy flux is a power law in frequency, $\nu F_\nu \propto \nu^s$, with spectral slope $s=(3-p)/2$ \citep[][]{rybicki_lightman_79}.

The {\it ansatz} of isotropy for the velocity distribution of the synchrotron-emitting particles is based on the assumption that the underlying acceleration mechanism does not imprint strong anisotropies, or that some other process (e.g., plasma instabilities) is capable of isotropizing the distribution on timescales shorter than the particle cooling time \citep[][]{Kulsrud2005,Longair2011}. In this work, we revisit these commonly adopted assumptions by means of first-principles particle-in-cell (PIC) simulations of particle acceleration in magnetically dominated plasma turbulence.

Recent studies of kinetic turbulence in highly magnetized pair plasmas have shown that nonthermal particle acceleration is a generic by-product of the turbulent energy cascade \citep{Zhdankin17,Zhdankin18,ComissoSironi18, ComissoSironi19, Nattila2019arXiv,Wong2020}. In particular, PIC simulations in unprecedentedly large domains have demonstrated that magnetic reconnection within the turbulent cascade is responsible for the initial particle energization, while large-scale turbulent fluctuations control the acceleration to higher energies \citep{ComissoSironi18,ComissoSironi19}. A crucial outcome of this energization mechanism is the generation of strongly anisotropic particle distributions \citep{ComissoSironi19}. Hence, one would expect the synchrotron emission of the turbulence-accelerated particles to deviate significantly from the standard expectation of an isotropic particle population.

In this paper, we quantify the energy dependence of particle anisotropy in magnetically dominated pair plasma turbulence, and we investigate its implications for the synchrotron emission. We show that the self-consistent generation of anisotropic particle distributions gives rise to hard (i.e., $s> 0.5$) synchrotron spectra across an extended frequency range even when the underlying particle spectra are relatively soft (i.e., $p> 2$). Remarkably, the synchrotron slope is nearly insensitive to the degree of initial turbulent fluctuations: weaker turbulence levels produce a stronger anisotropy, thereby counteracting the effect of the softer particle spectrum. Our findings may help explain the origin of the hard radio spectra of pulsar wind nebulae without the need to invoke hard ($p<2$) particle distributions.

\section{Analytical estimates}

We provide a simple analytical estimate of the synchrotron spectrum emitted by an anisotropic population of nonthermal particles \citep[see also Appendix A of][]{TavecchioSobacchi2020}. We assume that the number of particles per unit Lorentz factor is
\begin{equation}
\label{eq:Ngamma}
\frac{dN}{d\gamma}\propto \gamma^{-p} \quad{\rm for}\quad \gamma_{\min}<\gamma<\gamma_{\max}\;,
\end{equation}
and that the pitch angle (i.e., the angle $\alpha$ between the particle velocity and the local magnetic field) depends on the particle Lorentz factor as
\begin{equation}
\label{eq:alphagamma}
\sin\alpha=
\begin{cases}
\left(\frac{\gamma}{\gamma_{\rm crit}}\right)^q & {\rm for} \quad \gamma_{\min}<\gamma<\gamma_{\rm crit} \\ 
1 & {\rm for} \quad \gamma_{\rm crit}<\gamma<\gamma_{\max}
\end{cases}
\end{equation}
with $q>0$. This assumption is consistent with the results of first-principles PIC simulations of magnetically dominated turbulent plasmas \citep[][]{ComissoSironi19}, where it was shown that the velocity of low energy particles is preferentially aligned with the local magnetic field, while the most energetic particles are preferentially oriented in the plane perpendicular to the field. The origin of the anisotropy is linked to the acceleration mechanism that dominates for particles of a given energy. Particles at the low energy end of the nonthermal tail ($\gamma_{\rm min}<\gamma<\gamma_{\rm crit}$, with $\gamma_{\rm crit}\sim 10\,\gamma_{\min}$) are primarily accelerated by non-ideal electric fields aligned with the local magnetic field during reconnection, while the acceleration to larger Lorentz factors ($\gamma_{\rm crit}<\gamma<\gamma_{\rm max}$) is controlled by scattering off turbulent fluctuations, and therefore by electric fields perpendicular to the local magnetic field \citep{ComissoSironi18, ComissoSironi19}.


We characterize the synchrotron spectrum through the energy flux, $\nu F_\nu$. For the sake of simplicity, we assume that the magnetic field is a constant, $B_0$. For electrons with a Lorentz factor $\gamma<\gamma_{\rm crit}$, the emitted synchrotron frequency is $\nu \sim \gamma^2 \nu_L \sin\alpha \propto \gamma^{2+q}$, where $\nu_L= eB_0/2 \pi mc$ is the nonrelativistic Larmor frequency. The photon energy flux is equal to the number of electrons at a given $\gamma$, $N_\gamma \sim\gamma (dN/d\gamma) \propto \gamma^{1-p}$, multiplied by the power radiated by one electron, $P_{\rm sync} = 2 \sigma_{\rm T}c (B_0^2/8\pi)\gamma^2\sin^2\alpha\propto\gamma^{2+2q}$, with $\sigma_{\rm T}$ indicating the Thomson cross section. So when $\gamma<\gamma_{\rm crit}$, we have $\nu F_\nu \sim N_\gamma P_{\rm sync} \propto \gamma^{3-p+2q}\propto\nu^{(3-p+2q)/(2+q)}$, and we eventually find that
\begin{equation}
\label{eq:L}
\nu F_\nu \propto \nu^{\frac{3-p}{2}}\times
\begin{cases}
\left(\frac{\nu}{\nu_{\rm crit}}\right)^\frac{q(1+p)}{2(2+q)} & {\rm for} \quad \nu_{\min}<\nu<\nu_{\rm crit} \\
1 & {\rm for} \quad \nu_{\rm crit}<\nu<\nu_{\max}
\end{cases}
\end{equation}
where $\nu_{\min}\sim\gamma_{\min}^2\nu_L(\gamma_{\min}/\gamma_{\rm crit})^q$, $\nu_{\rm crit}\sim\gamma_{\rm crit}^2\nu_L$, and $\nu_{\max}\sim\gamma_{\max}^2\nu_L$. Note that $\nu_{\min}$ is lower by a factor of $(\gamma_{\min}/\gamma_{\rm crit})^q$ than in the isotropic case $q=0$.

An anisotropic pitch angle distribution can significantly harden the synchrotron spectrum. Assuming for example that $q=p-1$, Eq. \eqref{eq:L} gives $\nu F_\nu\propto\nu$ for $\nu_{\min}<\nu<\nu_{\rm crit}$, and the usual $\nu F_\nu\propto\nu^{(3-p)/2}$ for $\nu_{\rm crit}<\nu<\nu_{\max}$. Since typically $p\geq 2$, the energy flux for $\nu_{\rm crit} < \nu< \nu_{\max}$ scales as $\nu F_\nu\propto\nu^{1/2}$ or softer. The fact that $p$ and $q$ are correlated is clearly seen in PIC simulations (see below).

\section{Numerical Method and Setup}

In order to study the development of nonthermal particle acceleration and the associated synchrotron spectrum from magnetically dominated plasma turbulence, we perform \emph{ab initio} PIC simulations employing the PIC code TRISTAN-MP \citep{buneman_93,spitkovsky_05}. 
We conduct a suite of large-scale two-dimensional simulations (2D) and one large-scale three-dimensional (3D) simulation. In 2D our computational domain is a square of size $L^2$, while in 3D it is a cube of size $L^3$. We use periodic boundary conditions in all directions. For both 2D and 3D domains, all three components of particle momenta and electromagnetic fields are evolved in time. The effect of synchrotron cooling on the particle motion is neglected in all the simulations, which is valid as long as the cooling time of the highest energy particles is much longer than the lifetime of the system.

The simulation setup is similar to our previous works on magnetically dominated plasma turbulence \citep{ComissoSironi18,ComissoSironi19}. We initialize a uniform electron-positron plasma with total particle density $n_0$ according to a Maxwell-J\"{u}ttner distribution $f(\gamma) \propto n_0 \sqrt{\gamma^2 - 1}  \, e^{-\gamma/\theta_0}$ with thermal spread $\theta_0 = {k_B T_0}/{m c^2} = 0.3$, where $k_B$ indicates the Boltzmann constant and $T_0$ is the initial plasma temperature. The corresponding mean particle Lorentz factor at the initial time is $\gamma_{th0}\simeq1.58$. Turbulence develops from uncorrelated magnetic field fluctuations that are initialized in the plane perpendicular to a uniform mean magnetic field, which is taken along the $z$-direction, $B_0 {\bm{\hat z}}$. The initial fluctuations have low wavenumbers $k_j = 2\pi n_j/L$, with $n_j \in \{ {1, \ldots ,4} \}$ and $j$ indicating the wavenumber direction, and equal amplitude per mode. With this choice, the initial magnetic energy spectrum peaks near $k_p = 8 \pi /L$, which defines the energy-carrying scale $l = 2 \pi /k_p$ used as our unit length.

The strength of the initial magnetic field fluctuations is parameterized by the magnetization $\sigma_0  = \delta B_{{\rm{rms}}0}^2/4\pi h_0$, where  $\delta B_{{\rm{rms}}0} = \langle {\delta {B^2} (t=0)} \rangle^{1/2}$ is the space-averaged root-mean-square value of the initial magnetic field fluctuations and $h_0$ is the initial enthalpy density. Here we focus on the magnetically dominated regime, corresponding to $\sigma_0 \gg 1$. We use a fiducial magnetization $\sigma_0=75$. Our results do not depend on the choice of the initial thermal spread $\theta_0$, apart from an overall energy rescaling \citep[see][]{ComissoSironi19}.

The large size of our computational domain (with $L$ up to $32640$ cells in 2D and $2460$ cells in 3D) allows us to achieve asymptotically-converged results. We resolve the initial plasma skin depth $d_{e0}=c/\omega_{p0}=\sqrt {\gamma_{th0} {m}c^2/4\pi n_0 {e^2}}$ with 2 cells in 2D and 1.5 cells in 3D. The reference simulation (see Figs. \ref{fig1} and \ref{fig2}) has 64 computational particles per cell on average, while other simulations employ 16 particles per cell, as we have verified that this still properly captures the particles spectrum and anisotropy. The simulation time step is controlled by the numerical speed of light of 0.45 cells per time step.

\section{Results}

\begin{figure}
\begin{center}
\includegraphics[width=8.65cm]{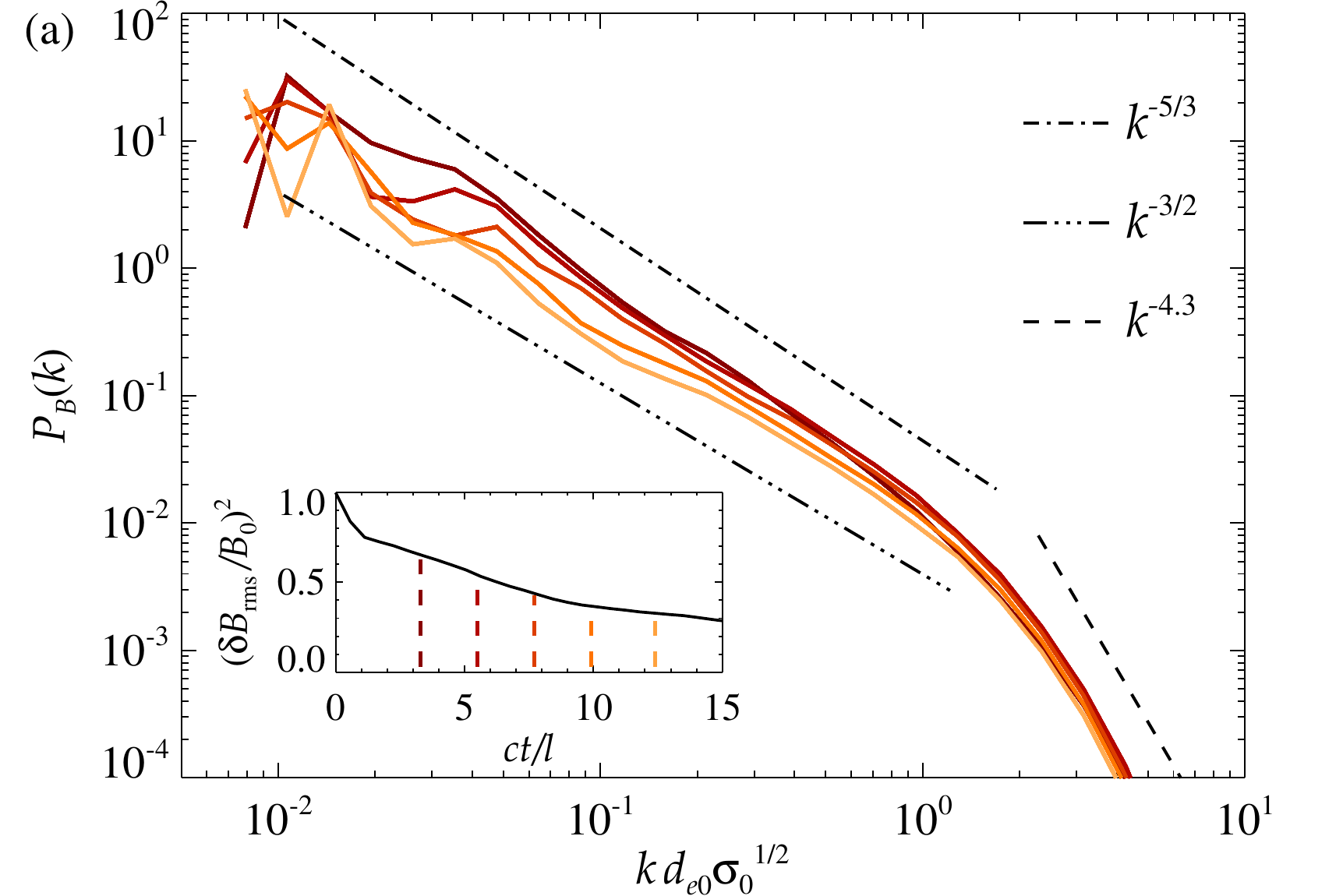}
\hspace*{0.215cm}\includegraphics[width=8.65cm]{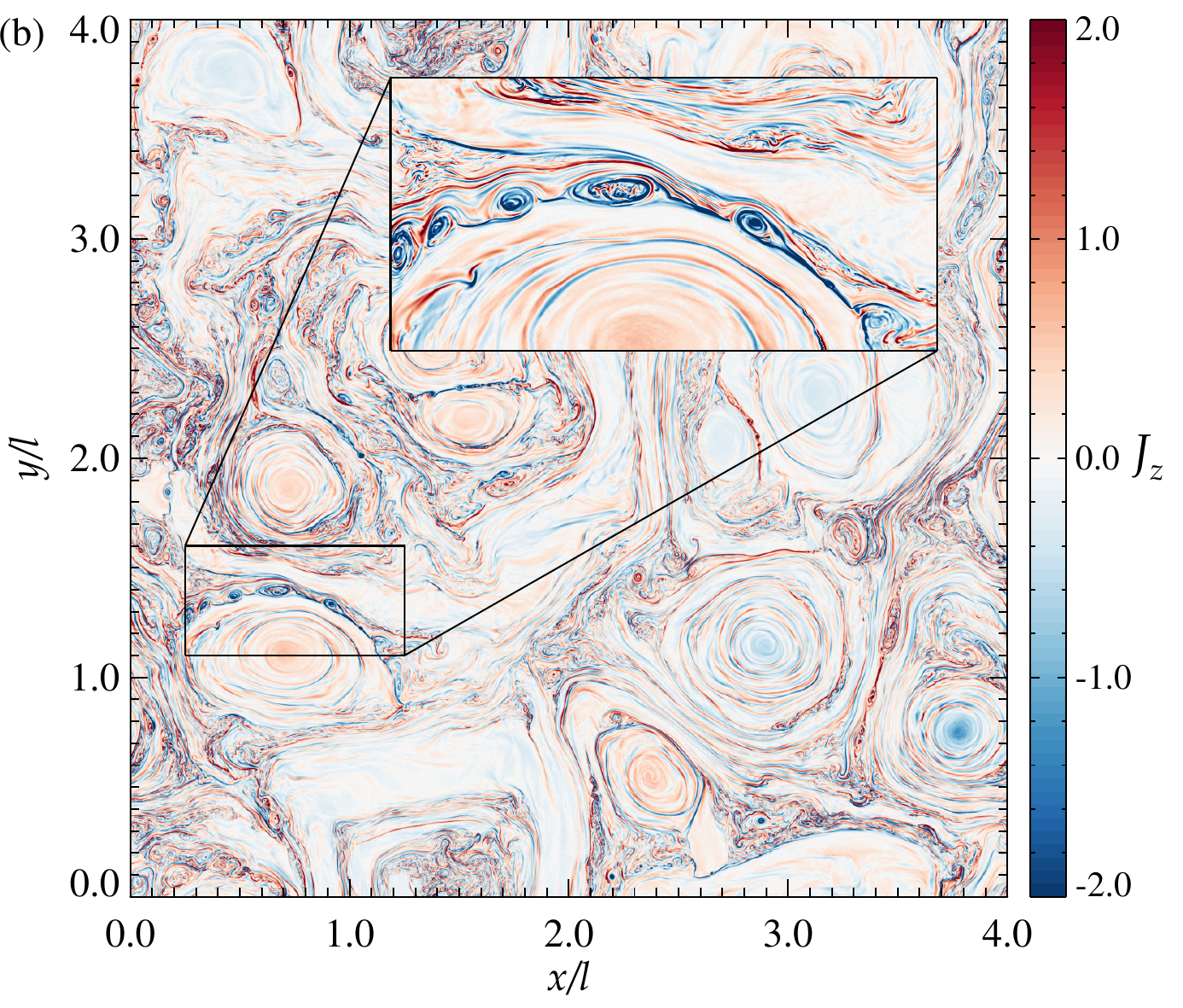}
\end{center}
\caption{Development of turbulence from a simulation with $\sigma_0=75$, $\delta B_{{\rm{rms}}0}/ B_0=1$, and $L/d_{e0}=8160$ (with $l=L/4$). Top panel: power spectrum of the magnetic field computed at different times, as indicated by the vertical dashed lines (same color coding) in the inset, which shows the time evolution of $\delta B_{\rm rms}^2=\langle {\delta {B^2}}\rangle$ normalized to $B_0^2$. Bottom panel: out-of-plane current density $J_z$ at $ct/l=4.4$ (normalized to $e n_0 c$) indicating the presence of current sheets and reconnection plasmoids (see inset).}
\label{fig1}
\end{figure}

In our simulations, the magnetic energy decays in time, as no continuous driving is imposed, and a well-developed inertial range and kinetic range of the turbulence cascade develop within the outer-scale nonlinear timescale. Fig.~\ref{fig1}(a) shows the time evolution of the magnetic power spectrum $P_B(k)$ from our reference simulation. Each curve refers to a different time (from brown to orange), as indicated by the corresponding vertical dashed lines in the inset, where we present the temporal decay of the energy in turbulent fluctuations $\delta B^2_{\rm rms}/B_0^2$. At late times, when most of the turbulent magnetic energy has been transferred to the particles, {the particle energy density is roughly in equipartition with the total field energy density.} The mean particle Lorentz factor increases to $\langle \gamma\rangle\sim \gamma_\sigma= (1+\sigma_0/2) \gamma_{th0}$, and the typical plasma skin depth becomes $d_e \sim d_{e0} \sigma_0^{1/2}$ for $\sigma_0 \gg 1$. From Fig. \ref{fig1}(a), we can see that at MHD scales ($k d_{e0} \sigma_0^{1/2} \lesssim 1$) the magnetic power spectrum is consistent with a Kolmogorov scaling $P_B(k) \propto k^{-5/3}$ \citep{Biskamp2003}, while the Iroshnikov-Kraichnan scaling $P_B(k) \propto k^{-3/2}$ \citep{Iroshnikov63,Kraichnan65} is possibly approached at late times. At kinetic scales, the spectrum steepens and approaches $P_B(k) \propto k^{-4.3}$ \citep{ComissoSironi19}. 

As shown in Fig.~\ref{fig1}(b), the turbulent cascade leads to the formation of intense current layers. Many of these layers become prone to fast magnetic reconnection due to the plasmoid instability that kicks in when the layers exceed a critical aspect ratio \citep{comisso_16,Comisso_ApJ2017, uzdensky_16}. Magnetic reconnection plays a crucial role in extracting particles from the thermal pool and injecting them into the acceleration process \citep{ComissoSironi18,ComissoSironi19}. This process leaves a signature in the orientation of the particle velocity with respect to the magnetic field, which, as we show below, has important consequences for the emitted synchrotron radiation.

\begin{figure}
\begin{center}
\includegraphics[width=8.75cm]{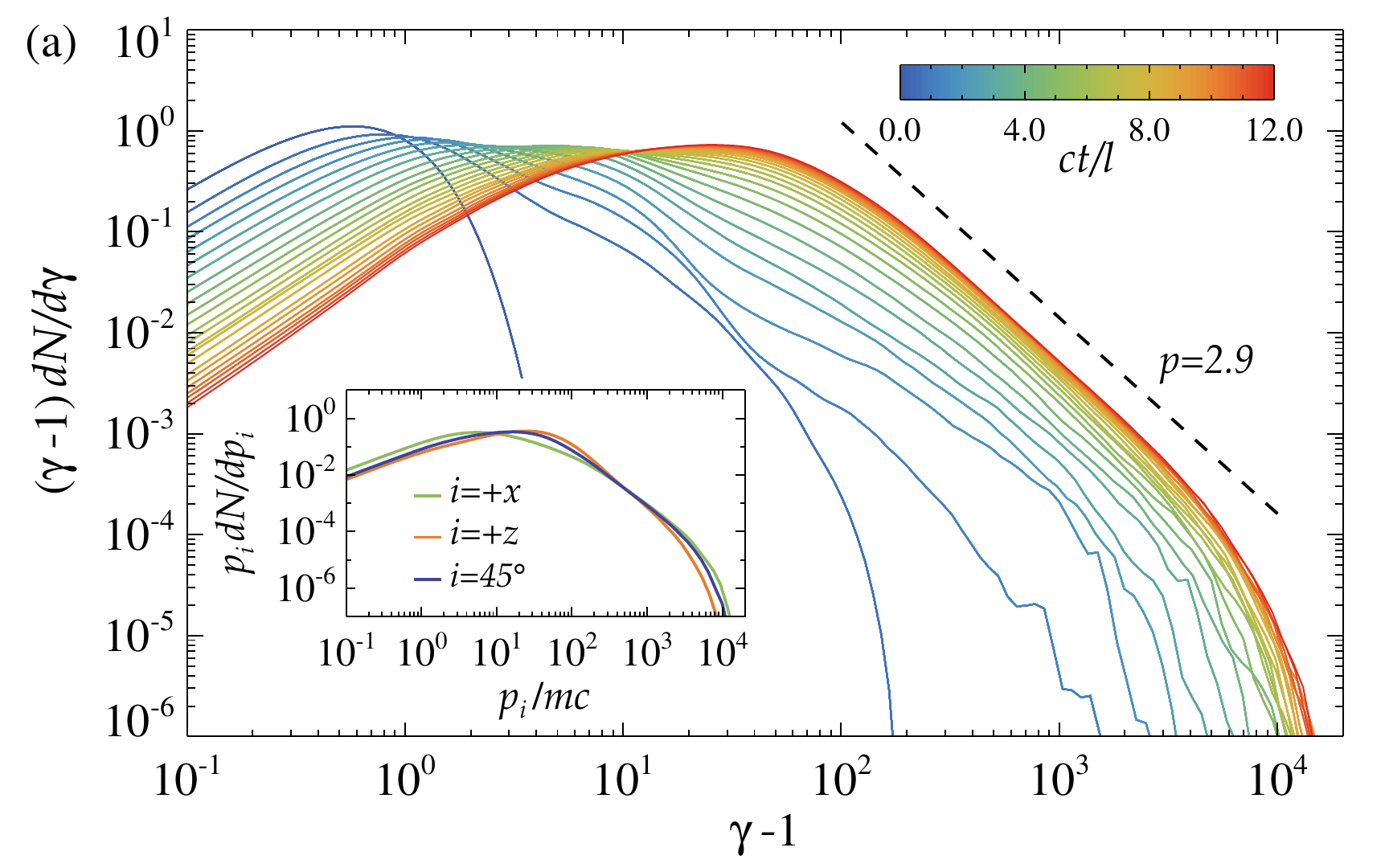}
\includegraphics[width=8.75cm]{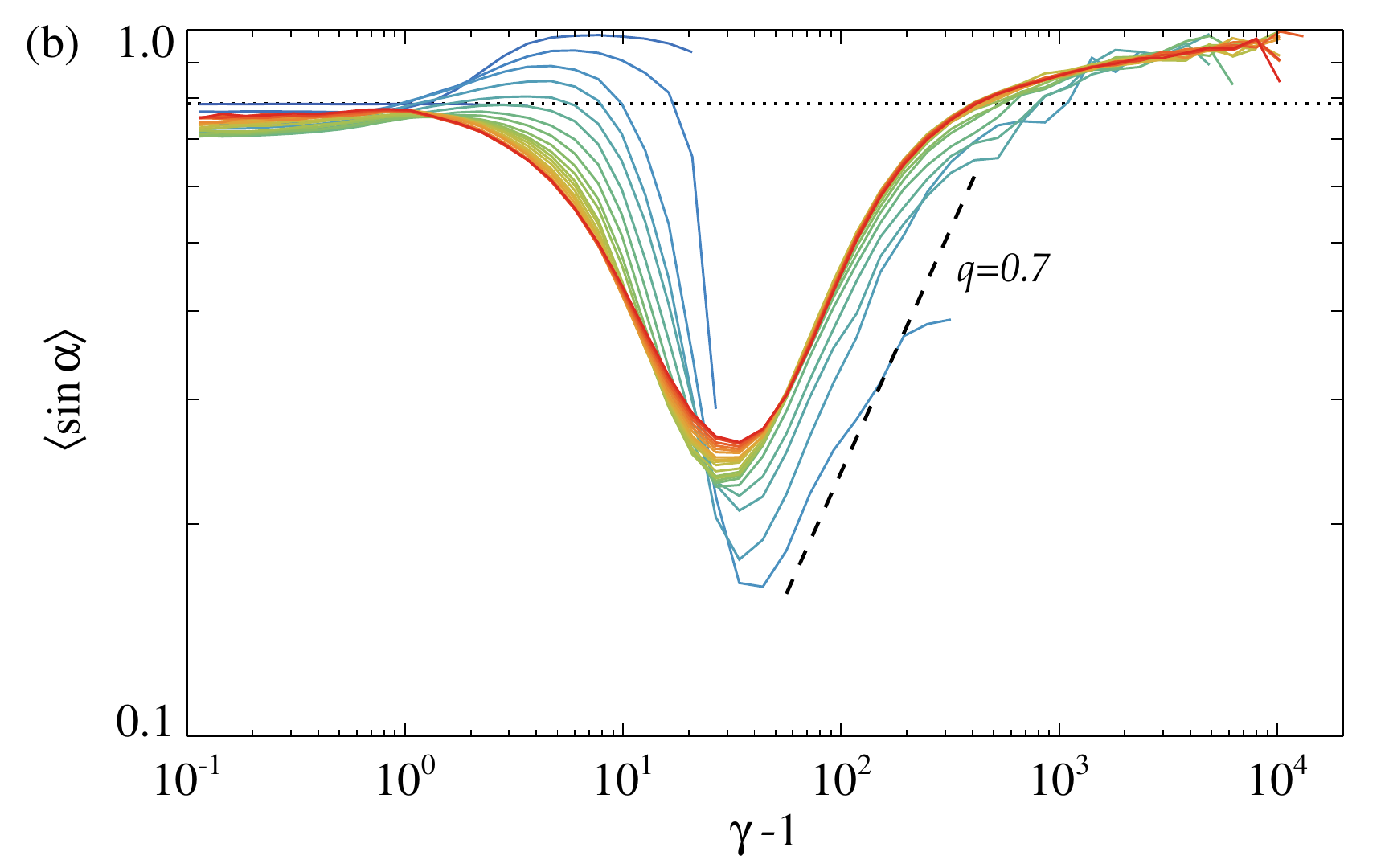}
\includegraphics[width=8.75cm]{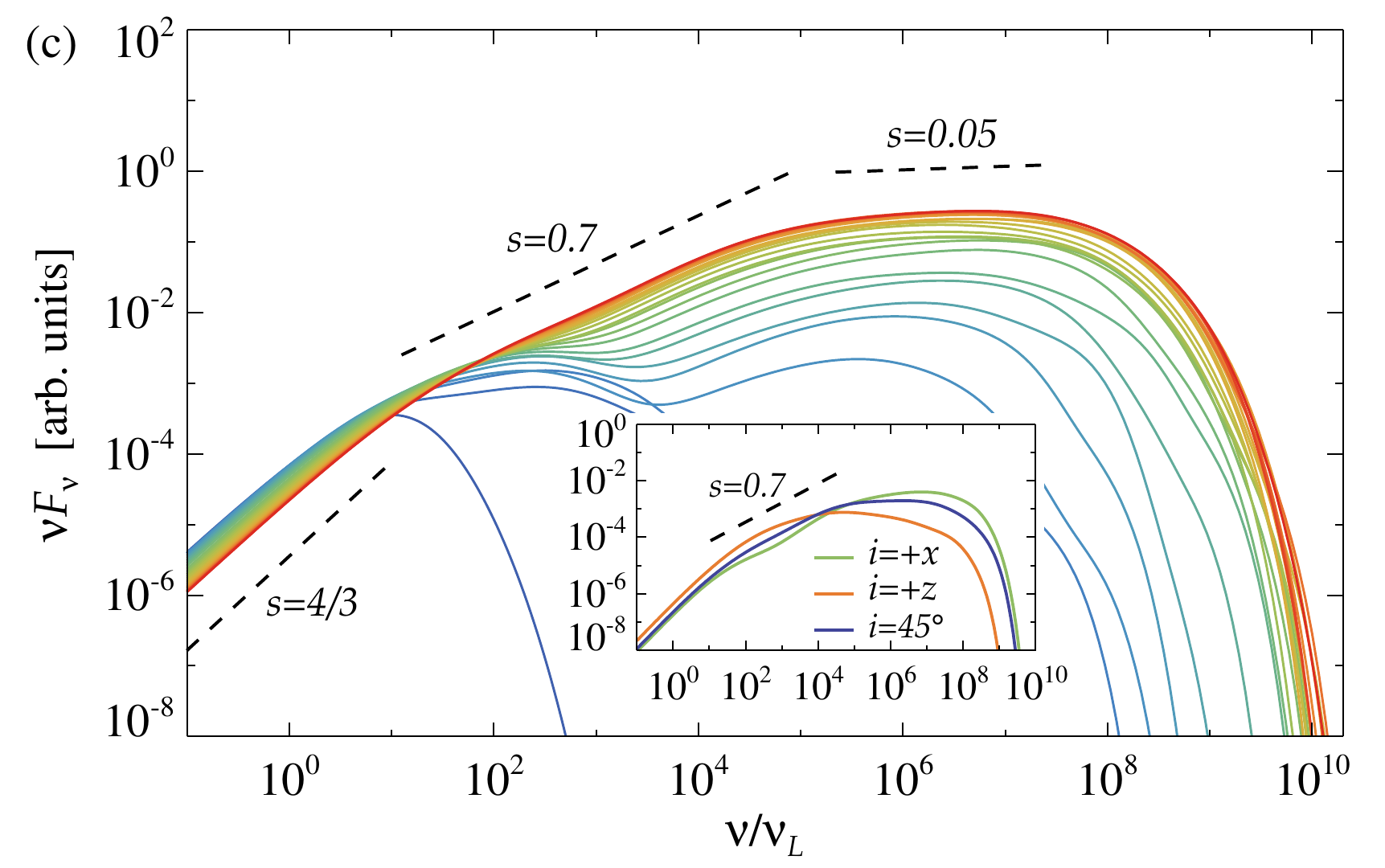}
\end{center}
\caption{Time evolution of the particle spectrum (top panel), mean pitch angle sine (middle panel), and angle-integrated synchrotron spectrum (bottom panel) for the simulation in Fig.~\ref{fig1}. The insets of the top and bottom panels show, respectively, the particle momentum spectra and the synchrotron spectra at late times ($ct/l=11$) along $+{\bm{\hat{x}}}$ (green line), $+{\bm{\hat{z}}}$ (orange line), and at $45$ degrees between $+{\bm{\hat{x}}}$ and $+{\bm{\hat{z}}}$ (blue line). }
\label{fig2}
\end{figure}  

In Fig.~\ref{fig2}(a), we show the time evolution of the particle spectrum $dN/d\gamma$. As a result of field dissipation, the spectrum shifts to energies much larger than the initial thermal energy. At late times, when most of the turbulent energy has decayed, the particle energy spectrum stops evolving (orange and red lines), and it peaks at $\gamma \sim 30$. It extends well beyond the peak into a nonthermal tail that can be described by a power law with slope $p = -d\log N/d\log (\gamma-1) \simeq 2.9$. From the inset in Fig.~\ref{fig2}(a), we can also see that the momenta of low energy particles are mostly in the direction of the mean magnetic field, while at higher energies, well into the nonthermal tail, particle momenta become mostly perpendicular to it. As we showed in \citet{ComissoSironi18,ComissoSironi19}, this is a consequence of the particle acceleration mechanism: electric fields aligned with the local magnetic field are important at low energies (typical Lorentz factors up to a few times higher than $\sigma_0 \gamma_{th0}$), while electric fields perpendicular to the magnetic field take over at higher energies.

The process of particle acceleration drives a significant energy-dependent anisotropy of the particle pitch angle $\alpha$, i.e. the angle between the particle velocity and the local magnetic field (see Figs. 19-21 in \citealt{ComissoSironi19}). In particular, in Fig.~\ref{fig2}(b) we show the average of $\sin \alpha$ as a function of $\gamma$. We measure $\alpha$ in the local ${\bm{E}} \times {\bm{B}}$ frame, since this frame is the appropriate one to compute the synchrotron emission. The measured mean deviates significantly from the expected mean for an isotropic distribution, $\langle \sin \alpha \rangle = \pi/4$ (compare with the dotted line). After a few outer-scale eddy turnover times, $\langle \sin \alpha \rangle$ settles to a steady state. The steady-state curve of $\langle \sin \alpha \rangle$ attains a minimum near the Lorentz factor corresponding to the peak of the particle spectrum at $\gamma \sim \gamma_{th0} \sigma_0/2 \sim 30$, and increases monotonically for higher energies up to $\langle \sin \alpha \rangle \simeq 1$. In particular, in the low energy range of the nonthermal tail, $40 \lesssim \gamma \lesssim 300$, the mean pitch angle sine follows an approximate power law $\langle \sin \alpha \rangle \propto \gamma^q$ with $q \simeq 0.7$. This implies that a particle with $\gamma\sim \gamma_{th0} \sigma_0/2$ will have a significantly weaker synchrotron emission than a particle in the high energy end of the nonthermal tail.

\begin{figure}
\includegraphics[width=8.75cm]{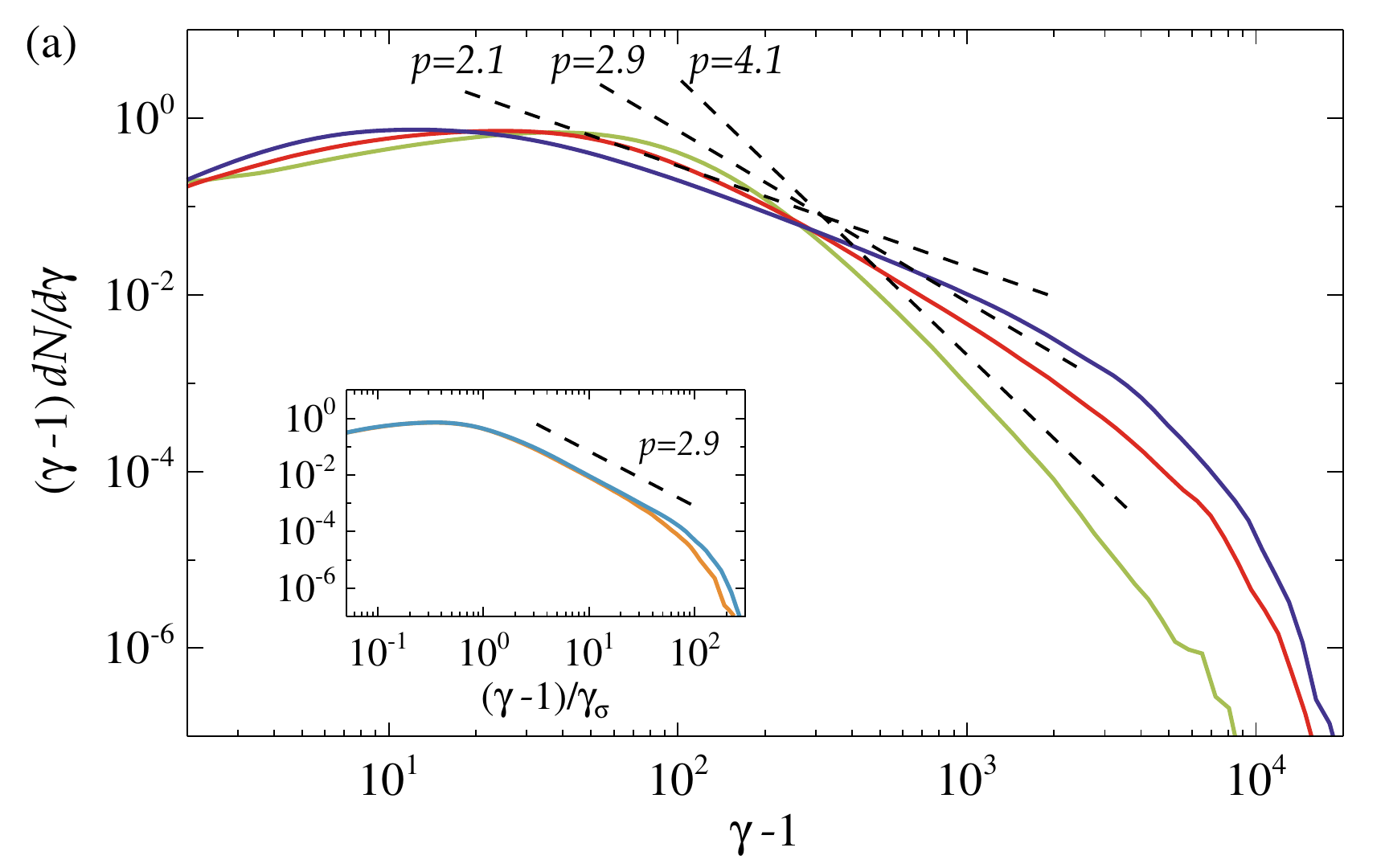}
\includegraphics[width=8.75cm]{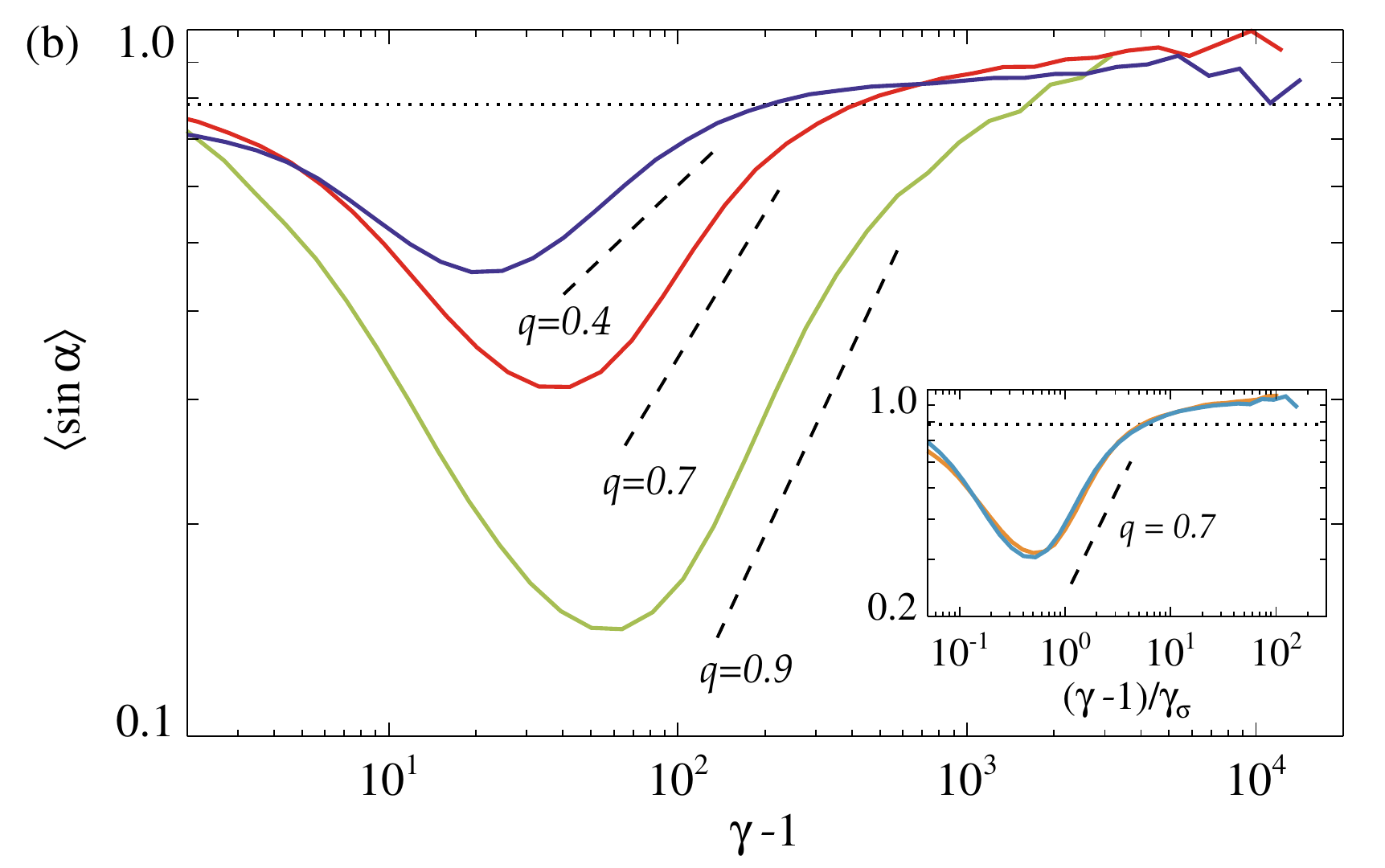}
\includegraphics[width=8.75cm]{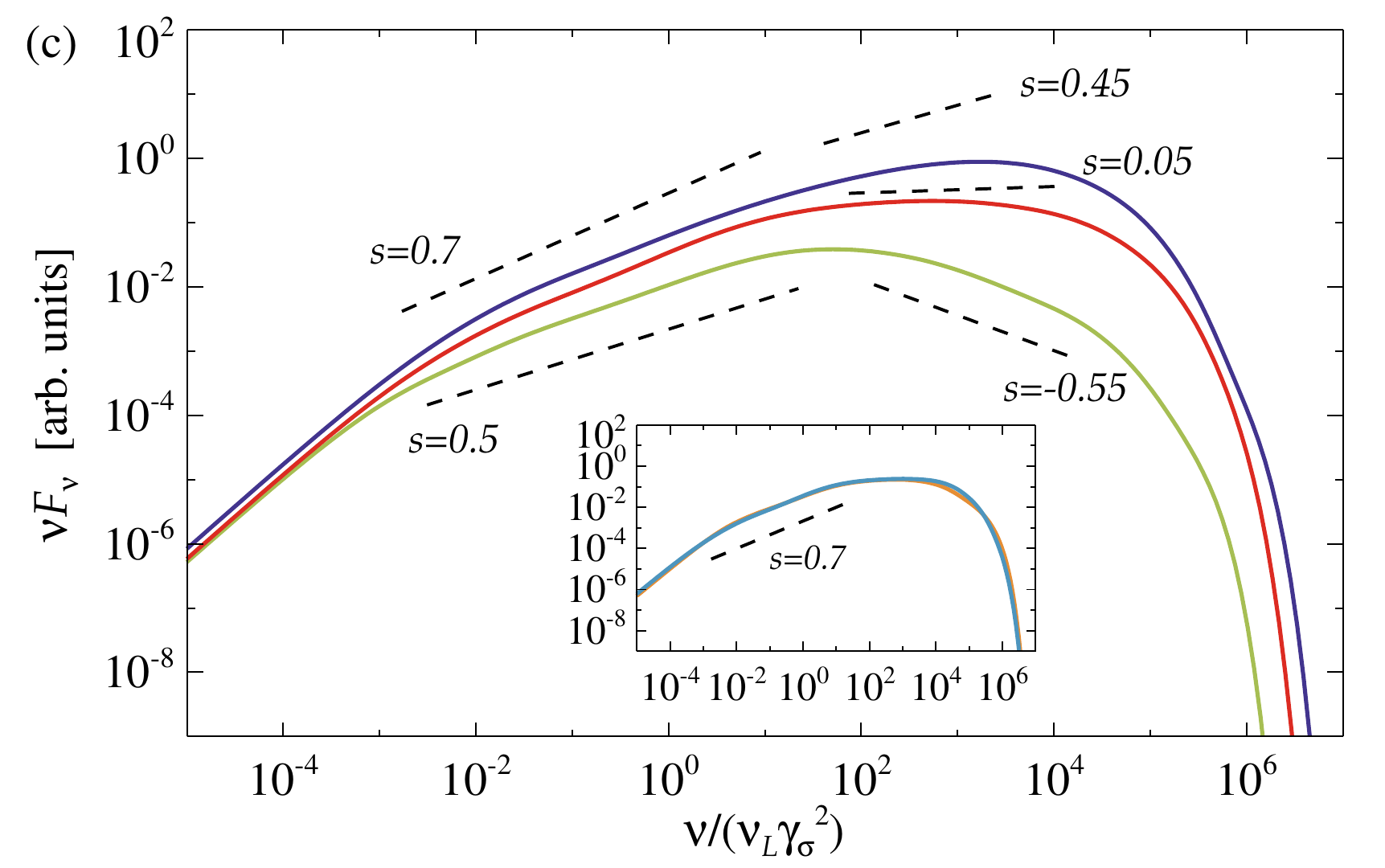}
\caption{Panels as in Fig.~\ref{fig2}, for cases with different $\delta B_{\rm rms0}^2/B_0^2=0.25$ (green), $1$ (red), and $4$ (blue) at late times ($ct/l=11$). 
We fix $\sigma_0=75$ and $L/d_{e0}=8160$.  The insets show the results from two additional simulations with $\delta B_{\rm rms0}^2/B_0^2 = 1$, $L/d_{e0}=8160$, and different magnetization $\sigma_0=75/2$ (orange line) and $\sigma_0=2 \times 75$ (cyan line).}
\label{fig3}
\end{figure}

We calculate the synchrotron spectrum by summing over the angle-integrated synchrotron emission from every particle in an incoherent way \citep{rybicki_lightman_79, RevilleKirk2010}. Fig.~\ref{fig2}(c) shows the time evolution of the angle-integrated synchrotron spectrum $\nu F_\nu$.\footnote{We normalize the synchrotron energy flux $\nu F_\nu$ to $P_{\gamma_\sigma} N_p$, where $P_{\gamma_\sigma}=(4/3)c\sigma_{\rm T} \gamma_\sigma^2 B_0^2/8 \pi$ is the mean synchrotron power per particle of an isotropic population with $\gamma=\gamma_\sigma$, while $N_p$ is the total number of particles.}  At very low frequencies, $\nu \ll  \gamma_\sigma^2 \nu_L \langle\sin\alpha\rangle_{\gamma_\sigma}$, where $\gamma_\sigma=(1+\sigma_0/2) \gamma_{th0}$ is the post-dissipation mean Lorentz factor and $\langle\sin\alpha\rangle_{\gamma_\sigma}$ is the mean pitch angle sine at $\gamma_\sigma$, the spectrum is the usual $\nu F_\nu \propto \nu^{4/3}$. At very high frequencies, the synchrotron spectrum is $\nu F_\nu \propto \nu^{({3-p})/{2}} = \nu^{0.05}$, which is consistent with the standard synchrotron spectral slope produced by an isotropic nonthermal particle population with $p=2.9$. However, there is an  intermediate frequency range --- extending over nearly four orders of magnitude in frequency ---  with a harder spectral slope, $\nu F_\nu \propto \nu^{0.7}$, which corresponds to the range where the emitting particles are anisotropic. Indeed, we have verified that the synchrotron spectral slope in this range, $s = 0.7$, is in line with the estimate from Eq. \eqref{eq:L}.

In the inset of Fig.~\ref{fig2}(c) we show the synchrotron spectrum at late times along different directions, by considering only the particles whose velocity vector falls within a solid angle $\Omega/4\pi=0.01$ around the line of sight. As expected, the anisotropy of the emitting particles leads to an anisotropic synchrotron emissivity. In the plane perpendicular to the background magnetic field (i.e., along the $+{\bm{\hat{x}}}$ direction), the synchrotron spectral slope is even harder than $s=0.7$. This is due to the fact that (i) high energy particles, which have $\langle\sin\alpha\rangle\sim 1$, emit preferentially in the direction perpendicular to the background field;\footnote{At late times, turbulent magnetic fluctuations have decayed, so the direction of the local magnetic field in the ${\bm{E}} \times {\bm{B}}$ frame, where the pitch angle should be computed, nearly coincides with the $+{\bm{\hat{z}}}$ direction of the background field.} (ii)  low energy particles, which have $\langle\sin\alpha\rangle\ll 1$, emit preferentially in the direction of the background field, and therefore do not contribute to the synchrotron emission along $+{\bm{\hat{x}}}$. For a similar reason, the synchrotron spectrum in the  $+{\bm{\hat{z}}}$ direction of the background field is softer than $s=0.7$.

In order to understand the dependence of the synchrotron spectrum on the main physical parameters that govern the problem, we perform a suite of simulations with different physical conditions. The results from simulations with different values of $\delta B_{\rm rms0}^2/B_0^2 = 0.25, 1, 4$ are shown in the main panels of Fig.~\ref{fig3}. A striking result is that in the anisotropy dominated range, the synchrotron spectra $\nu F_\nu$ have very similar slopes (see Fig. \ref{fig3}(c)), despite significant differences in the slopes of the particle spectra in Fig.~\ref{fig3}(a). This is due to the fact that the anisotropy is stronger (i.e., $\langle\sin\alpha\rangle$ reaches lower values) for a weaker level of turbulent fluctuations. Indeed, when the mean magnetic field is stronger, i.e. $\delta B_{\rm rms}^2/B_0^2$ is smaller, the rate of pitch angle isotropization is lower since the pitch angle scattering timescale is $t_{\rm scatt} \propto B_0^2/\delta B_{\rm rms}^2$, which allows the accelerated particles to approach the strongest anisotropy attainable in reconnection. At higher energies, scattering off turbulent fluctuations drives the emitting particles in the direction perpendicular to the local magnetic field \citep{ComissoSironi18, ComissoSironi19}, and the synchrotron spectrum approaches the expected scaling $\nu F_\nu \propto \nu^{({3-p})/{2}}$. As shown in the insets, where we investigate the dependence of our results on $\sigma_0$, for large values of the magnetization  the differences in the particle spectrum and the pitch angle anisotropy are minor, which leads to a synchrotron spectrum with a slope of $s \simeq 0.7$ for both $\sigma_0=75/2$ and $\sigma_0= 2 \times 75$.

\begin{figure}
\begin{center}
\includegraphics[width=8.75cm]{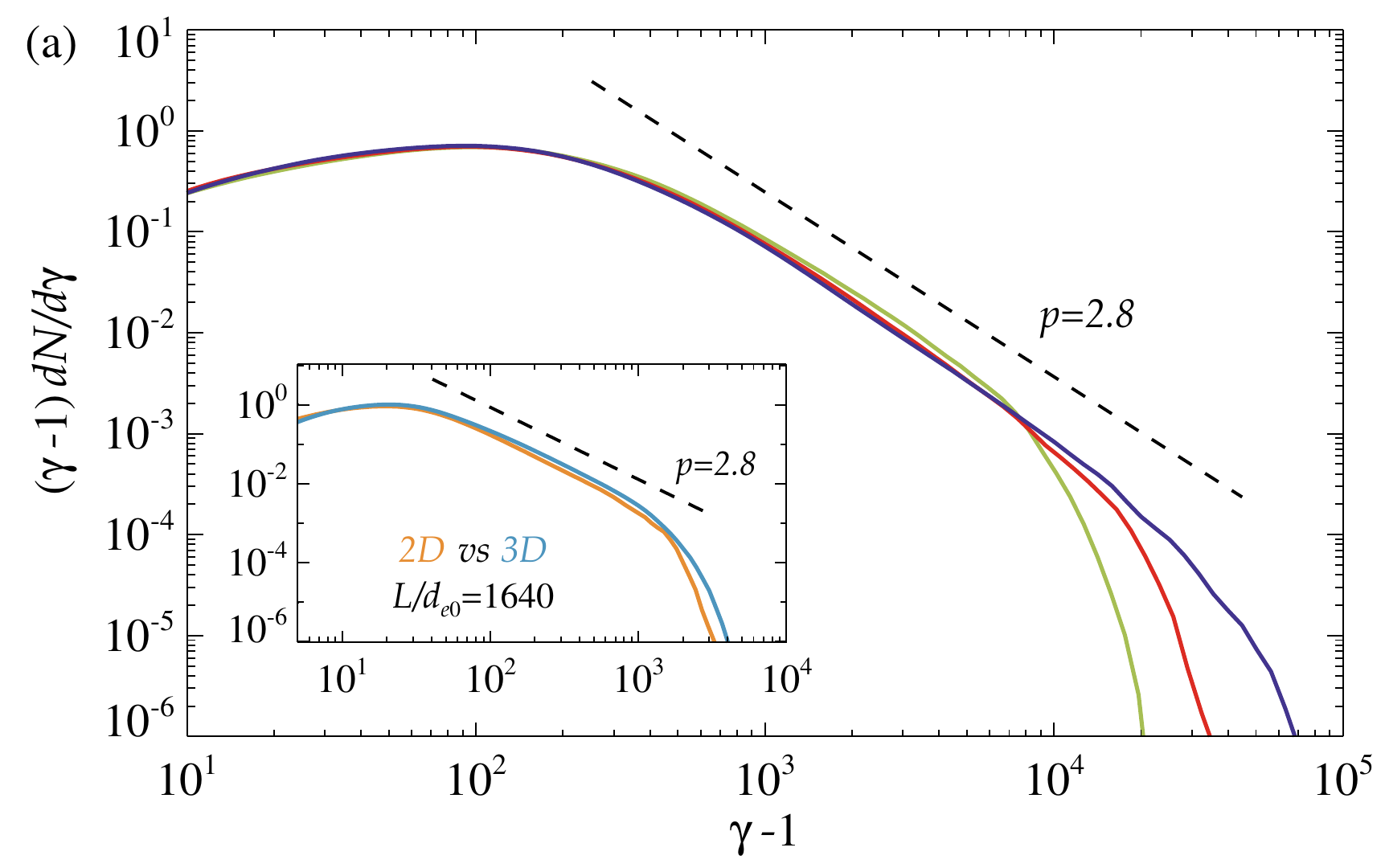}
\includegraphics[width=8.75cm]{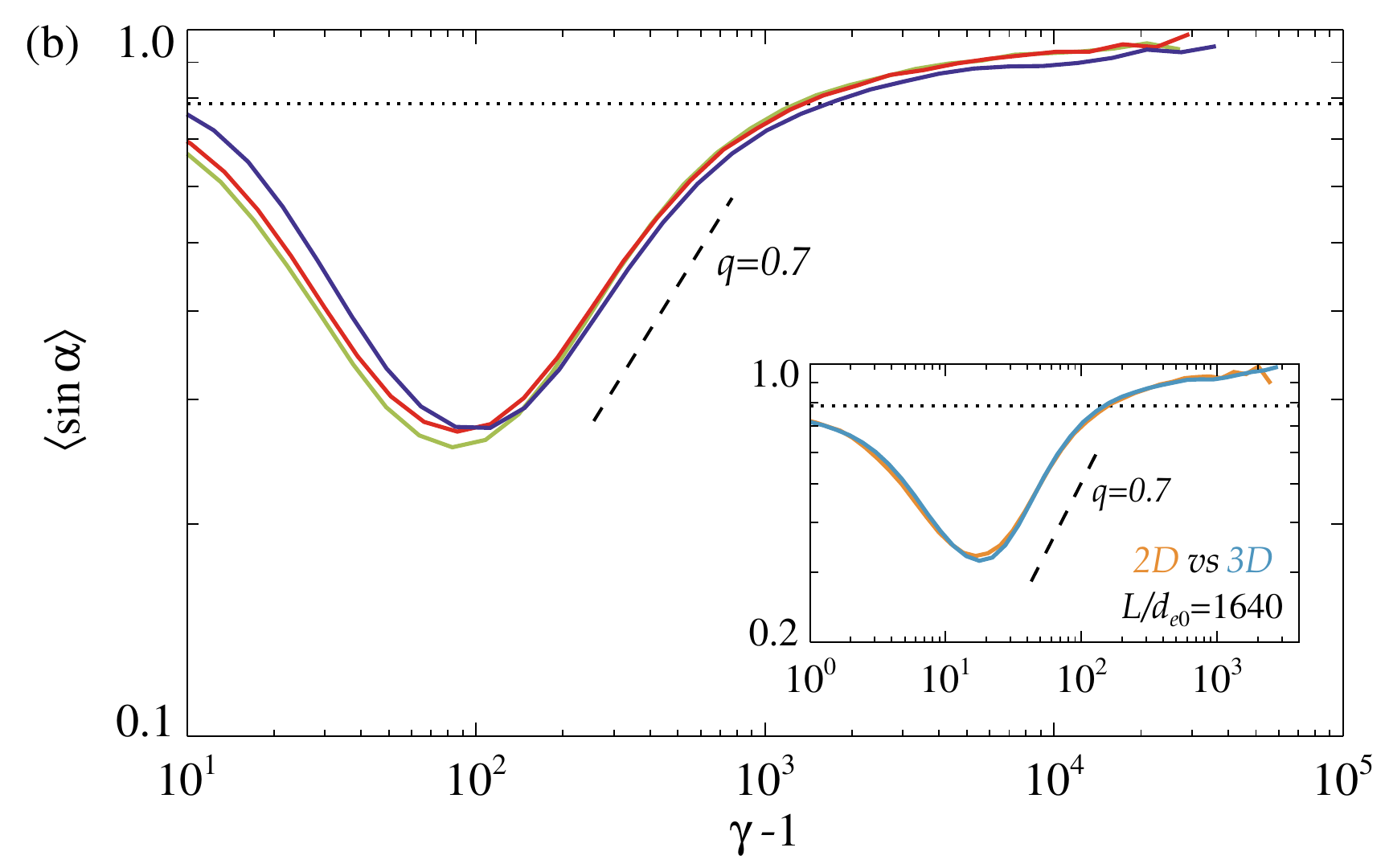}
\includegraphics[width=8.75cm]{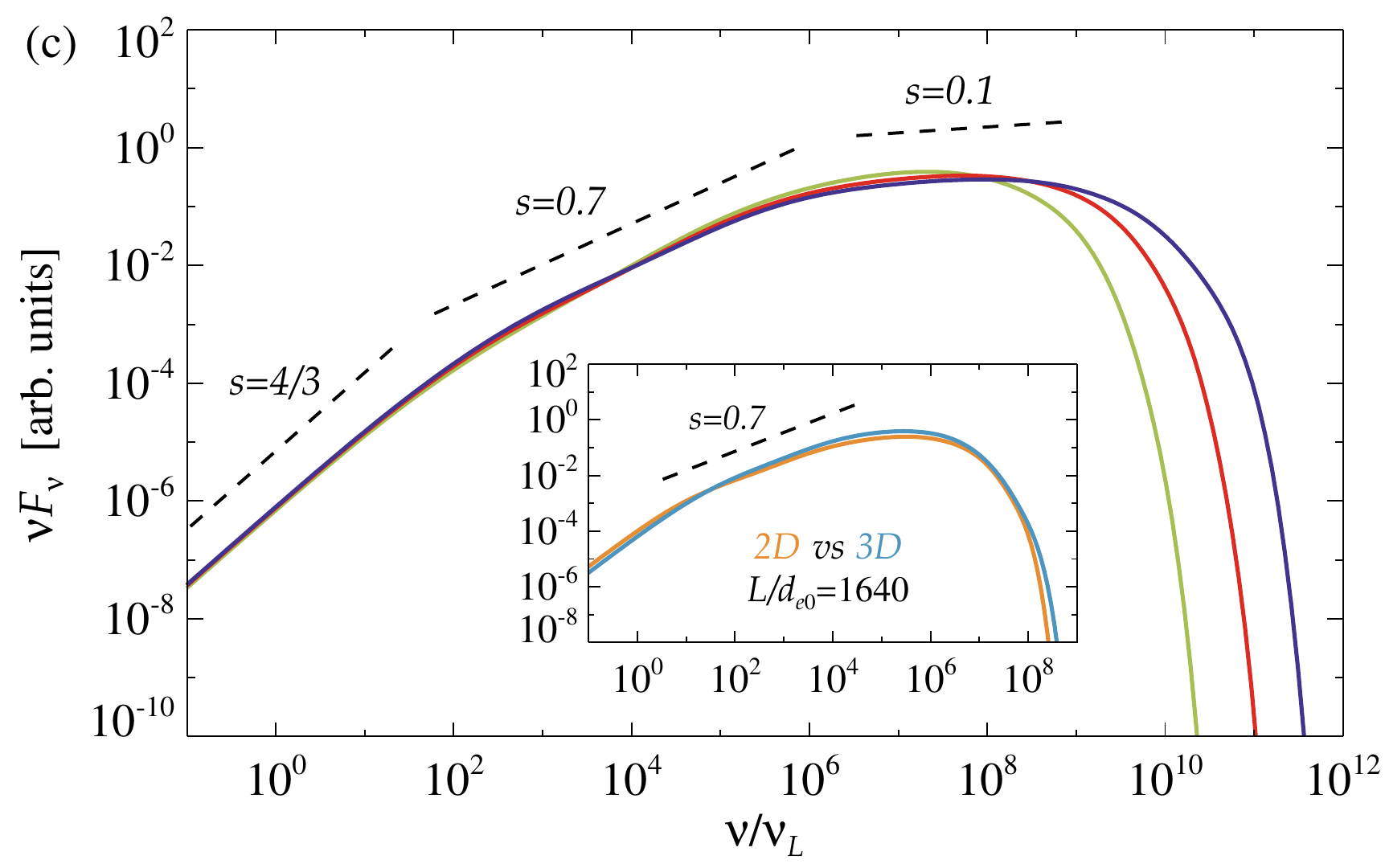}
\end{center}
\caption{Panels as in Fig.~\ref{fig2}, for cases with different system size $L/d_{e0}=4080$ (green), $8160$ (red), and $16320$ (blue)  at late times ($ct/l=11$).
We fix $\delta B_{{\rm{rms}}0}/ B_0 = 1$ and $\sigma_0=300$. The insets show the comparison between 3D (cyan line) and 2D (orange line) simulations with $\delta B_{{\rm{rms}}0}/ B_0 = 1$, $\sigma_0=40$, and $L/d_{e0}=1640$, taken at late times when the residual turbulent magnetic energy in the two simulations is about the same.}
\label{fig4}
\end{figure}

In Fig.~\ref{fig4} we show the dependence of our results on the system size. As can be seen from Fig.~\ref{fig4}(a), the particle spectra at late times from simulations with different system size differ only in the extension of the high energy tail, which extends up to the cutoff Lorentz factor $\gamma_{\max} \sim{\sqrt{\sigma_z}\gamma_{th0}(l/d_{e0})}$, where $\sigma_z=B_0^2/4\pi h_0=\sigma_0(B_0^2/\delta B_{{\rm{rms}}0}^2)$ is the magnetization associated with the mean magnetic field \citep{ComissoSironi18}. Fig.~\ref{fig4}(b) shows that the pitch angle anisotropy remains essentially the same for different domain sizes. This translates directly to the synchrotron spectrum shown in Fig.~\ref{fig4}(c), which exhibits the same slope of $s \simeq 0.7$ in the anisotropy dominated frequency range, while the high-frequency cutoff increases quadratically with the system size since $\nu_{\max} \sim \gamma_{\max}^2 \nu_L$.

We have shown in \citet{ComissoSironi18,ComissoSironi19} that the basic ingredients of the synchrotron spectrum, namely the particle spectrum and the characteristic pitch angle, are remarkably similar between 2D and 3D simulations. As a further confirmation, in the insets of Figs. \ref{fig4}(a) and \ref{fig4}(b) we show the particle spectrum and the mean pitch angle sine at late times from 3D and 2D simulations with the same physical parameters, which display a remarkable agreement. As a consequence, the synchrotron spectrum computed from the 3D simulation, which is shown in the inset of Fig. \ref{fig4}(c) (cyan line), is very close to the one computed from the corresponding 2D simulation (orange line). Again, the angle-integrated synchrotron spectrum displays a slope $s \simeq 0.7$ in the anisotropy dominated frequency range.

\section{Discussion}

We have shown that the angle-integrated synchrotron spectrum of electrons accelerated by magnetically dominated turbulence can be remarkably hard, $\nu F_\nu\propto\nu^s$ with $s\sim 0.5-0.7$, over a range of nearly four orders of magnitude in frequency. The spectral slope may be even harder, $s\gtrsim 0.7$, if the observer is in the plane perpendicular to the mean magnetic field. The spectral hardness does not arise because the particle energy spectrum itself is hard, i.e., $dN/d\gamma \propto \gamma^{-p}$ with $p<2$, but rather because the distribution is anisotropic. Within the nonthermal tail of accelerated particles, lower energy electrons tend to be more aligned with the local magnetic field than higher energy electrons, so their synchrotron emission is comparatively weaker. This energy-dependent anisotropy has the effect of hardening the synchrotron spectrum.\footnote{We expect similar conclusions to hold for electrons accelerated by reconnection with a strong guide field, even in a non-turbulent environment. Indeed, in our simulations of magnetically dominated turbulence, the acceleration of particles attaining the strongest anisotropy is primarily due to reconnection \citep{ComissoSironi19}. In laminar reconnection, the guide field would play the role of the mean field of our turbulent setup.}

Our results have important implications for the interpretation of the hard radio spectra of Pulsar Wind Nebulae (PWNe), where the observed radio spectral index is typically $s\sim 0.7-1$ \citep{GaenslerSlane2006, Reynolds+2017}. In the notable case of the Crab Nebula, the radio spectral index is $s\sim 0.7$ \citep{Hester2008, Buhler2014}. Turbulence, which develops as an effect of MHD instabilities in PWNe \citep{Porth2014MNRAS}, has been suggested to accelerate the radio-emitting electrons \citep{LyutikovMNRAS2019, Xu+2019}. However, these studies simply invoked a hard particle spectrum ($p<2$) over some energy range, without addressing from first principles the origin of the accelerated particles. We have shown that producing hard radio spectra does not necessarily require an acceleration mechanism that transfers most of the available energy to a small fraction of particles (which is the case when $p<2$). Instead, hard synchrotron spectra may be produced by a relatively soft population of nonthermal particles with an energy-dependent anisotropy. Finally, note that our assumption of neglecting the effect of synchrotron cooling on the particle evolution is well justified in PWNe, where the cooling time of  radio-emitting electrons is longer than the age of the system.

For the Crab Nebula, it is generally quoted that the hard spectral slope persists over six orders of magnitude in frequency (for an extended discussion, see e.g. \citealt{bietenholz_97}). Since the hard range of our synchrotron spectra extends over four decades, contributions from turbulent regions with moderately different properties would be required. While a detailed comparison to the Crab Nebula radio emission is left for future work, we mention that in our model we would imply that (i) the total number of radio-emitting electrons is at least one order of magnitude higher than in the case with isotropic electrons. More precisely, the number of radio electrons increases by a factor of $(\nu_{\rm crit}/\nu_{\min})^{{q}/({2+q})}$; (ii) the initial magnetization of the plasma is relatively low and/or the pre-dissipation plasma is relatively cold, $\gamma_\sigma \sqrt{\langle\sin\alpha\rangle_{\gamma_\sigma}} \lesssim 10^3$, where $\gamma_\sigma\sim \gamma_{th0}\sigma_0$ and $\langle\sin\alpha\rangle_{\gamma_\sigma}$ is the mean pitch angle of the electrons at $\gamma\sim\gamma_\sigma$. This constraint comes from the fact that in the Crab $\nu_{\min}\sim \gamma_\sigma^2\nu_L \langle\sin\alpha\rangle_{\gamma_\sigma}\lesssim \rm{GHz}$, and assuming a magnetic field of $B_0\sim 0.2 \,\rm{mG}$.

Anisotropic distributions of the emitting electrons --- which we have demonstrated to be a natural by-product of magnetically dominated turbulence \citep{ComissoSironi19} --- have been invoked in other astrophysical sources, including Gamma Ray Bursts (GRBs) and blazars. In GRBs, anisotropic distributions may explain (i) low-frequency spectra harder than the synchrotron line of death $\nu F_\nu \propto \nu^{4/3}$ in the prompt emission \citep{LloydPetrosian2000, LloydPetrosian2002, YangZhang2018}; (ii) fast variability \citep{Beloborodov+2011} and some degree of circular polarization (\citealt{Wiersema2014}; see also \citealt{Nava+2016}) in the afterglow. Some of these models require the electron pitch angle to be $\lesssim 1/\gamma$, which might be possible if the electrons are nonrelativistic before the turbulent component of the magnetic field is dissipated. In blazars, the magnetization of the emission region that is inferred from the spectral modeling may significantly increase if the electron distribution is anisotropic, which is more consistent with the expected magnetic nature of jets \citep{SobacchiLyubarsky2019, SobacchiLyubarsky2020, TavecchioSobacchi2020}. Considering the effect of synchrotron and/or inverse Compton cooling on the electron distribution, as required to model GRBs and blazars, is left for a future study.

\acknowledgments

We acknowledge fruitful discussions with Dimitrios Giannios, Daniel Groselj, Yuri Lyubarsky, Maxim Lyutikov, and Fabrizio Tavecchio. This research acknowledges support from the Sloan Fellowship in Physics, DoE DE-SC0016542, NASA ATP NNX17AG21G, and NSF/AAG AST-1716567. The simulations were performed on Columbia University (Habanero), NASA-HEC (Pleiades), and NERSC (Cori) resources. \\

\bibliographystyle{aasjournal}
\bibliography{Synchrotron_Turbulence}

\begin{thebibliography}{}
\expandafter\ifx\csname natexlab\endcsname\relax\def\natexlab#1{#1}\fi
\providecommand{\url}[1]{\href{#1}{#1}}
\providecommand{\dodoi}[1]{doi:~\href{http://doi.org/#1}{\nolinkurl{#1}}}
\providecommand{\doeprint}[1]{\href{http://ascl.net/#1}{\nolinkurl{http://ascl.net/#1}}}
\providecommand{\doarXiv}[1]{\href{https://arxiv.org/abs/#1}{\nolinkurl{https://arxiv.org/abs/#1}}}

\bibitem[{{Beloborodov} {et~al.}(2011){Beloborodov}, {Daigne}, {Mochkovitch},
  \& {Uhm}}]{Beloborodov+2011}
{Beloborodov}, A.~M., {Daigne}, F., {Mochkovitch}, R., \& {Uhm}, Z.~L. 2011,
  \mnras, 410, 2422, \dodoi{10.1111/j.1365-2966.2010.17616.x}

\bibitem[{{Bietenholz} {et~al.}(1997){Bietenholz}, {Kassim}, {Frail}, {Perley},
  {Erickson}, \& {Hajian}}]{bietenholz_97}
{Bietenholz}, M.~F., {Kassim}, N., {Frail}, D.~A., {et~al.} 1997, \apj, 490,
  291, \dodoi{10.1086/304853}

\bibitem[{{Biskamp}(2003)}]{Biskamp2003}
{Biskamp}, D. 2003, {Magnetohydrodynamic Turbulence} (Cambridge University
  Press)

\bibitem[{{B{\"u}hler} \& {Blandford}(2014)}]{Buhler2014}
{B{\"u}hler}, R., \& {Blandford}, R. 2014, Reports on Progress in Physics, 77,
  066901, \dodoi{10.1088/0034-4885/77/6/066901}

\bibitem[{{Buneman}(1993)}]{buneman_93}
{Buneman}, O. 1993, {in ``Computer Space Plasma Physics'', Terra Scientific,
  Tokyo, 67}

\bibitem[{{Comisso} {et~al.}(2016){Comisso}, {Lingam}, {Huang}, \&
  {Bhattacharjee}}]{comisso_16}
{Comisso}, L., {Lingam}, M., {Huang}, Y.-M., \& {Bhattacharjee}, A. 2016,
  Physics of Plasmas, 23, 100702, \dodoi{10.1063/1.4964481}

\bibitem[{{Comisso} {et~al.}(2017){Comisso}, {Lingam}, {Huang}, \&
  {Bhattacharjee}}]{Comisso_ApJ2017}
{Comisso}, L., {Lingam}, M., {Huang}, Y.~M., \& {Bhattacharjee}, A. 2017, \apj,
  850, 142, \dodoi{10.3847/1538-4357/aa9789}

\bibitem[{{Comisso} \& {Sironi}(2018)}]{ComissoSironi18}
{Comisso}, L., \& {Sironi}, L. 2018, \prl, 121, 255101,
  \dodoi{10.1103/PhysRevLett.121.255101}

\bibitem[{{Comisso} \& {Sironi}(2019)}]{ComissoSironi19}
---. 2019, \apj, 886, 122, \dodoi{10.3847/1538-4357/ab4c33}

\bibitem[{{Gaensler} \& {Slane}(2006)}]{GaenslerSlane2006}
{Gaensler}, B.~M., \& {Slane}, P.~O. 2006, \araa, 44, 17,
  \dodoi{10.1146/annurev.astro.44.051905.092528}

\bibitem[{{Hester}(2008)}]{Hester2008}
{Hester}, J.~J. 2008, \araa, 46, 127,
  \dodoi{10.1146/annurev.astro.45.051806.110608}

\bibitem[{{Iroshnikov}(1963)}]{Iroshnikov63}
{Iroshnikov}, P.~S. 1963, \azh, 40, 742

\bibitem[{{Kraichnan}(1965)}]{Kraichnan65}
{Kraichnan}, R.~H. 1965, Physics of Fluids, 8, 1385, \dodoi{10.1063/1.1761412}

\bibitem[{{Kulsrud}(2005)}]{Kulsrud2005}
{Kulsrud}, R.~M. 2005, {Plasma physics for astrophysics} (Princeton University
  Press)

\bibitem[{{Lloyd} \& {Petrosian}(2000)}]{LloydPetrosian2000}
{Lloyd}, N.~M., \& {Petrosian}, V. 2000, \apj, 543, 722, \dodoi{10.1086/317125}

\bibitem[{{Lloyd-Ronning} \& {Petrosian}(2002)}]{LloydPetrosian2002}
{Lloyd-Ronning}, N.~M., \& {Petrosian}, V. 2002, \apj, 565, 182,
  \dodoi{10.1086/324484}

\bibitem[{{Longair}(2011)}]{Longair2011}
{Longair}, M.~S. 2011, {High Energy Astrophysics} (Cambridge University Press)

\bibitem[{{Lyutikov} {et~al.}(2019){Lyutikov}, {Temim}, {Komissarov}, {Slane},
  {Sironi}, \& {Comisso}}]{LyutikovMNRAS2019}
{Lyutikov}, M., {Temim}, T., {Komissarov}, S., {et~al.} 2019, \mnras, 489,
  2403, \dodoi{10.1093/mnras/stz2023}

\bibitem[{{N{\"a}ttil{\"a}}(2019)}]{Nattila2019arXiv}
{N{\"a}ttil{\"a}}, J. 2019, arXiv e-prints, arXiv:1906.06306.
\newblock \doarXiv{1906.06306}

\bibitem[{{Nava} {et~al.}(2016){Nava}, {Nakar}, \& {Piran}}]{Nava+2016}
{Nava}, L., {Nakar}, E., \& {Piran}, T. 2016, \mnras, 455, 1594,
  \dodoi{10.1093/mnras/stv2434}

\bibitem[{{Porth} {et~al.}(2014){Porth}, {Komissarov}, \&
  {Keppens}}]{Porth2014MNRAS}
{Porth}, O., {Komissarov}, S.~S., \& {Keppens}, R. 2014, \mnras, 438, 278,
  \dodoi{10.1093/mnras/stt2176}

\bibitem[{{Reville} \& {Kirk}(2010)}]{RevilleKirk2010}
{Reville}, B., \& {Kirk}, J.~G. 2010, \apj, 724, 1283,
  \dodoi{10.1088/0004-637X/724/2/1283}

\bibitem[{{Reynolds} {et~al.}(2017){Reynolds}, {Pavlov}, {Kargaltsev},
  {Klingler}, {Renaud}, \& {Mereghetti}}]{Reynolds+2017}
{Reynolds}, S.~P., {Pavlov}, G.~G., {Kargaltsev}, O., {et~al.} 2017, \ssr, 207,
  175, \dodoi{10.1007/s11214-017-0356-6}

\bibitem[{{Rybicki} \& {Lightman}(1979)}]{rybicki_lightman_79}
{Rybicki}, G.~B., \& {Lightman}, A.~D. 1979, {Radiative Processes in
  Astrophysics} (John Wiley \& Sons, Inc.)

\bibitem[{{Sobacchi} \& {Lyubarsky}(2019)}]{SobacchiLyubarsky2019}
{Sobacchi}, E., \& {Lyubarsky}, Y.~E. 2019, \mnras, 484, 1192,
  \dodoi{10.1093/mnras/stz044}

\bibitem[{{Sobacchi} \& {Lyubarsky}(2020)}]{SobacchiLyubarsky2020}
---. 2020, \mnras, 491, 3900, \dodoi{10.1093/mnras/stz3313}

\bibitem[{{Spitkovsky}(2005)}]{spitkovsky_05}
{Spitkovsky}, A. 2005, in AIP Conf. Ser., Vol. 801, Astrophysical Sources of
  High Energy Particles and Radiation, ed. {T.~Bulik, B.~Rudak, \&
  G.~Madejski}, 345

\bibitem[{{Tavecchio} \& {Sobacchi}(2020)}]{TavecchioSobacchi2020}
{Tavecchio}, F., \& {Sobacchi}, E. 2020, \mnras, 491, 2198,
  \dodoi{10.1093/mnras/stz3168}

\bibitem[{{Uzdensky} \& {Loureiro}(2016)}]{uzdensky_16}
{Uzdensky}, D.~A., \& {Loureiro}, N.~F. 2016, Physical Review Letters, 116,
  105003, \dodoi{10.1103/PhysRevLett.116.105003}

\bibitem[{{Wiersema} {et~al.}(2014){Wiersema}, {Covino}, {Toma}, {van der
  Horst}, {Varela}, {Min}, {Greiner}, {Starling}, {Tanvir}, {Wijers},
  {Campana}, {Curran}, {Fan}, {Fynbo}, {Gorosabel}, {Gomboc}, {G{\"o}tz},
  {Hjorth}, {Jin}, {Kobayashi}, {Kouveliotou}, {Mundell}, {O'Brien}, {Pian},
  {Rowlinson}, {Russell}, {Salvaterra}, {di Serego Alighieri}, {Tagliaferri},
  {Vergani}, {Elliott}, {Fari{\~n}a}, {Hartoog}, {Karjalainen}, {Klose},
  {Knust}, {Levan}, {Schady}, {Sudilovsky}, \& {Willingale}}]{Wiersema2014}
{Wiersema}, K., {Covino}, S., {Toma}, K., {et~al.} 2014, \nat, 509, 201,
  \dodoi{10.1038/nature13237}

\bibitem[{{Wong} {et~al.}(2020){Wong}, {Zhdankin}, {Uzdensky}, {Werner}, \&
  {Begelman}}]{Wong2020}
{Wong}, K., {Zhdankin}, V., {Uzdensky}, D.~A., {Werner}, G.~R., \& {Begelman},
  M.~C. 2020, \apjl, 893, L7, \dodoi{10.3847/2041-8213/ab8122}

\bibitem[{{Xu} {et~al.}(2019){Xu}, {Klingler}, {Kargaltsev}, \&
  {Zhang}}]{Xu+2019}
{Xu}, S., {Klingler}, N., {Kargaltsev}, O., \& {Zhang}, B. 2019, \apj, 872, 10,
  \dodoi{10.3847/1538-4357/aafb2e}

\bibitem[{{Yang} \& {Zhang}(2018)}]{YangZhang2018}
{Yang}, Y.-P., \& {Zhang}, B. 2018, \apjl, 864, L16,
  \dodoi{10.3847/2041-8213/aada4f}

\bibitem[{{Zhdankin} {et~al.}(2018){Zhdankin}, {Uzdensky}, {Werner}, \&
  {Begelman}}]{Zhdankin18}
{Zhdankin}, V., {Uzdensky}, D.~A., {Werner}, G.~R., \& {Begelman}, M.~C. 2018,
  \apj, 867, L18, \dodoi{10.3847/2041-8213/aae88c}

\bibitem[{{Zhdankin} {et~al.}(2017){Zhdankin}, {Werner}, {Uzdensky}, \&
  {Begelman}}]{Zhdankin17}
{Zhdankin}, V., {Werner}, G.~R., {Uzdensky}, D.~A., \& {Begelman}, M.~C. 2017,
  \prl, 118, 055103, \dodoi{10.1103/PhysRevLett.118.055103}

\end{thebibliography}

\end{document}